\documentclass[12pt]{article}
\usepackage[utf8]{inputenc}   
\usepackage{graphicx}         
\usepackage{cite}             
\usepackage{geometry}         
\usepackage[round]{natbib}

\usepackage{amsmath, amssymb, amsfonts}
\usepackage{graphicx,xcolor}
\usepackage{natbib}
\usepackage[linesnumbered,ruled,vlined]{algorithm2e}  
\usepackage{comment}
\usepackage{caption}
\usepackage{subcaption}
\usepackage{booktabs}
\usepackage{newtxtext}
\usepackage[subscriptcorrection]{newtxmath}
\usepackage{float}

\newcommand{\Real}{\mathbb{R}}
\newcommand{\vecop}{\operatorname{vec}}

\newcommand{\Xb}{\mathbf{X}}

\newcommand{\Yb}{\mathbf{Y}}
\newcommand{\yb}{\mathbf{y}}
\newcommand{\Zb}{\mathbf{Z}}
\newcommand{\zb}{\mathbf{z}}
\newcommand{\Cb}{\mathbf{C}}
\newcommand{\Ab}{\mathbf{A}}
\newcommand{\Bb}{\mathbf{B}}

\newcommand{\Cbbb}{\mathbb{C}}
\newcommand{\Xbbb}{\mathbb{X}}
\newcommand{\ab}{\mathbf{a}}
\newcommand{\bb}{\mathbf{b}}
\newcommand{\Ub}{\mathbf{U}}
\newcommand{\ub}{\mathbf{u}}
\newcommand{\Ib}{\mathbf{I}}

\newcommand{\Ob}{\mathbf{O}}

\newcommand{\Gb}{\mathbf{G}}

\newcommand{\mb}{\mathbf{m}}

\newcommand{\Cc}{{\cal C}}
\newcommand{\Bc}{{\cal B}}
\newcommand{\Ac}{{\cal A}}

\newcommand{\gammab}{\boldsymbol{\gamma}}
\newcommand{\Gammab}{\boldsymbol{\Gamma}}
\newcommand{\Sigmab}{\boldsymbol{\Sigma}}
\newcommand{\Omegab}{\boldsymbol{\Omega}}
\newcommand{\omegab}{\boldsymbol{\omega}}
\newcommand{\Deltab}{\boldsymbol{\Delta}}

\newcommand{\calX}{\mathcal{X}}
\newcommand{\calT}{\mathcal{T}}
\newcommand{\Psib}{\boldsymbol{\Psi}}

\newcommand{\calPG}{\mathcal{PG}}

\usepackage{amsmath, amssymb}

\newtheorem{theorem}{Theorem}[section]     
\newtheorem{definition}{Definition}[section] 

\title{A Bayesian Sparse Kronecker Product Decomposition Framework for Tensor Predictors with Mixed-Type Responses}

\author{
  \makebox[.45\linewidth][c]{%
    \parbox{.4\linewidth}{\centering
      Shao-Hsuan Wang\\
     Graduate Institute of Statistics, National Central University, Taoyuan, Taiwan\\
      \texttt{picowang@gmail.com}
    }
  }
  \hfill
  \makebox[.45\linewidth][c]{%
    \parbox{.4\linewidth}{\centering
      Hsin-Hsiung Huang\\
      Department of Statistics and Data Science, University of Central Florida, Orlando, Florida, U.S.A.\\
      \texttt{Hsin.Huang@ucf.edu}
    }
  }
}

\date{\today}  

\begin{document}

\maketitle

\begin{abstract}
Ultra-high-dimensional tensor predictors are increasingly common in neuroimaging and other biomedical studies, yet existing methods rarely integrate continuous, count, and binary responses in a single coherent model.  We present a \emph{Bayesian Sparse Kronecker Product Decomposition} (BSKPD) that represents each regression (or classification) coefficient tensor as a low-rank Kronecker product whose factors are endowed with element-wise Three-Parameter Beta–Normal shrinkage priors, yielding voxel-level sparsity and interpretability.  Embedding Gaussian, Poisson, and Bernoulli outcomes in a unified exponential-family form, and combining the shrinkage priors with P\'olya–Gamma data augmentation, gives closed-form Gibbs updates that scale to full-resolution 3-D images.  We prove posterior consistency and identifiability—even when each tensor mode dimension grows sub-exponentially with the sample size—thereby extending high-dimensional Bayesian theory to mixed-type multivariate responses.  Simulations and applications to ADNI and OASIS magnetic-resonance imaging datasets show that BSKPD delivers sharper signal recovery and lower predictive error than current low-rank or sparsity-only competitors while preserving scientific interpretability.
\end{abstract}

\vspace{1em}
\noindent\textbf{keywords:} ~
Bayesian tensor modeling; mixed-type multivariate outcomes; neuroimaging; P\'olya–Gamma augmentation; posterior contraction; sparse Kronecker product decomposition; Three-Parameter Beta–Normal prior

\newpage

\section{Introduction}\label{sec:intro}

Modern neuroimaging studies increasingly seek to relate high-resolution three-dimensional brain images to multiple clinical outcomes of varying types, including continuous ventricular volume, binary disease status, and ordinal dementia scores. Each image is naturally represented as a tensor comprising up to $10^7$ voxels, yielding a number of regression coefficients that may exceed the sample size by several orders of magnitude. Accordingly, a suitable method must simultaneously accommodate estimation, variable selection, and valid uncertainty quantification in an ultrahigh-dimensional tensor regression setting with mixed-type responses. However, existing approaches address only subsets of these challenges. For example,  \cite{guhaniyogi2017bayesian} propose a Bayesian tensor regression model with multiway shrinkage priors for scalar responses, yet do not accommodate response heterogeneity. Wu et al.  extend Bayesian tensor models to binary responses using tensor candecomp/parafac (CP) decomposition to reduce ultrahigh dimensional tensor with  P\'olya–Gamma augmentation, demonstrating strong performance in Alzheimer’s disease imaging studies. Nevertheless, their approach does not support mixed response types. Recent work by \cite{wang2024bayesian} develops a fully Bayesian tensor-on-tensor regression framework that achieves efficient computation and uncertainty quantification, though it remains limited to continuous outcomes. Classical approaches such as the generalized tensor regression model of \cite{zhou2013tensor}- preserve spatial structure and reduce dimensionality, yet lack full Bayesian inference and do not support heterogeneous outcomes. On the other hand, 
a multivariate Bayesian shrinkage framework of \cite{wang2025two} accommodates heterogeneous responses but operates on vectorized images, ignoring spatial structure. To the best of our knowledge, no existing methodology simultaneously satisfies all of the following: (i) support for mixed response types; (ii) preservation of tensor structure; (iii) computational scalability in ultrahigh-dimensional settings; and (iv) valid uncertainty quantification with theoretical guarantees. Addressing this gap is essential for principled and interpretable inference in modern neuroimaging applications. 

On the other hand,  \cite{wu2023sparse} introduced a 
sparse Kronecker-product decomposition (SKPD) approach, which preserves tensor geometry and offers substantial dimension reduction, yet existing implementations are largely limited to Gaussian responses and do not accommodate general outcome types. In this paper, we propose a Bayesian sparse Kronecker product decomposition (BSKPD) model, which embeds the sparse Kronecker-product decomposition of \cite{wu2023sparse} within the multivariate Bayesian hierarchical framework developed by \cite{wang2025two}.  To induce sparsity and mitigate overfitting in this high-dimensional regime, we adopt the three-parameter beta-normal (TPBN) priors of \cite{armagan2011generalized}, a flexible global-local shrinkage family that encompasses several well-known priors, including the horseshoe, Strawderman--Berger, and normal--exponential--gamma priors. Unlike spike-and-slab shrinkage approaches \citep{george1993variable,mitchell1988bayesian}, global-local shrinkage employs fully continuous priors that do not explicitly set coefficients to zero. This property makes it particularly suitable for factor models, imaging, signal processing.  Moreover, global-local priors are computationally more scalable and well-suited to variational inference or MCMC methods 
\citep{bhattacharya2015dirichlet,
carvalho2010horseshoe,
polson2010shrink}. The TPBN priors are governed by three parameters $(a, u, \tau)$, where  $a$ and $u$ control the tail behaviour and concentration of the local shrinkage through a beta-prime distribution, and $\tau$ regulates global shrinkage. Theoretical properties of TPBN priors have been established in high-dimensional settings \cite{bai2018high, wang2025two}. Notably,  \cite{wang2025two} introduced the MtMBSP method, which applies TPBN priors within a multivariate Bayesian variable selection framework to accommodate mixed-type outcomes. They
utilized an efficient Gibbs sampler via P\'olya–Gamma (PG) data augmentation \citep{polson2013bayesian}, which is a powerful technique for facilitating fully conjugate Bayesian inference in logistic and other non-Gaussian likelihood models. It is particularly useful in Bayesian logistic regression and related models involving logit or probit link functions, where standard Gibbs sampling is often intractable due to non-conjugacy. The key idea is to augment a latent variable drawn from the PG distribution that transforms the non-conjugate likelihood into a conditionally Gaussian form, allowing for efficient Gibbs sampling within a fully Bayesian framework. This approach enables scalable and exact posterior computation in high-dimensional settings and has been extended to various applications, including binary regression, negative binomial regression, and hierarchical models. Inherent to the BSKPD framework, we establish two main theoretical results: (a) an identifiability issue; and (b) posterior consistency, both in classical and high-dimensional settings. In a numerical study, we assessed the proposed method using two neuroimaging datasets: 416 baseline T1-weighted images from the OASIS-1 cross-sectional cohort \citep{mcevoy2010quantitative} and 246 scans from the ADNI-1 study \citep{jack2008alzheimer}. Adjusting for age and sex, the BSKPD model jointly predicts ventricular volume, Mini-Mental State Examination (MMSE) scores, and Clinical Dementia Ratings (CDR). It achieves predictive accuracy comparable to, and in some cases surpassing, that of black-box alternatives, while simultaneously identifying anatomically interpretable regions associated with neurodegeneration.

The paper is organized as follows. Section~\ref{sec:model} introduces the proposed model and computational framework. Theoretical results are presented in Section~\ref{sec:theory}, followed by empirical applications in Section~\ref{sec:experiments}. Concluding remarks are provided in Section~\ref{sec:discussion}.

\section{Bayesian SKPD}\label{sec:model}
In this section, we introduce the Bayesian modeling framework. First, we present a matrix-variate model with mixed-type responses. Second, we formulate the Bayesian model and discuss the identifiability issues concerning the coefficients associated with the covariates.

 \subsection{Tensor model with mixed-type responses} \label{sec:mim}
 We consider independent and identically distributed (i.i.d.) observations $\{(\mathbf{y}_i, {\cal X}_i, \mathbf{z}_i)\}_{i=1}^n$, where $\mathbf{y}_i = (y_{i1}, \dots, y_{iK})$ is a $K$-dimensional mixed-type response vector, ${\calX}_i \in \mathbb{R}^{D_1\times D_2\times D_3}$ is a third-order tensor of covariates (e.g., a 3D brain image), and $\mathbf{z}_i \in \mathbb{R}^q$ represents additional vector-valued predictors (e.g., age, demographic, or clinical variables in MRI studies). Let $\langle \cdot, \cdot \rangle$ denote the Frobenius inner product. For each observation $i$, we consider a mixed-type response regression model for $k = 1, \dots, K$. Specifically, if $y_{ik}$ is a binary response, we model it via a Logistic regression:
\begin{eqnarray}
\log\left( \frac{\Pr(y_{ik} = 1 \mid {\cal X}_i, \mathbf{z}_i)}{1 - \Pr(y_{ik} = 1 \mid{\cal X}_i, \mathbf{z}_i)} \right) = \langle {\cal X}_i, {\cal C}_k \rangle + \langle \mathbf{z}_i, \gamma_k \rangle + u_{ik};\label{logist}
\end{eqnarray}
otherwise, if $y_{ik}$ is continuous, it is modeled via a Gaussian regression:
\begin{eqnarray}
y_{ik} = \langle {\cal X}_i, {\cal C}_k \rangle + \langle \mathbf{z}_i, \gamma_k \rangle + \epsilon_{ik} + u_{ik},
\label{gaussian}
\end{eqnarray}
where $\Cc_k \in \mathbb{R}^{D_1 \times D_2 \times D_3}$ is the coefficient tensor associated with ${\cal X}_i$, and $\gamma_k \in \mathbb{R}^q$ is the coefficient vector modeling the effects of the auxiliary covariates $\mathbf{z}_i$, facilitating interpretability. Furthermore, all $\{\epsilon_{ik}\}$ represent independent and identically distributed (i.i.d.) error terms, where $\epsilon_{ik} \sim N(0, 1)$. The values $\{ u_{iK}\}$s capture the latent correlation structure among the responses and is assumed to follow a multivariate normal distribution $N_K(\mathbf{0}, \boldsymbol{\Sigma})$, where $\boldsymbol{\Sigma}$ is a symmetric and positive definite covariance matrix.

For convenience, we omit the subscript $k$ from $\Cc_k$ and simply use $\Cc$ to represent the general form of the coefficient tensor throughout the remainder of this section. 
Denote $\otimes$ as the Kronecker product.  Given two tensors ${\Ac} \in \Real^{p_1\times p_2\times p_3}$ and ${\Bc} \in \Real^{q_1\times q_2\times q_3}$, the tensor Kronecker product of ${\Ac}$ and ${\Bc}$, still used as ${\Ac}\otimes {\Bc}$, is defined as 
\[
{\Ac} \otimes {\Bc}\in \Real^{
(p_1d_1)\times (p_2d_2)\times (p_3d_3)},~~
({\Ac} \otimes {\Bc})_{\cdot \cdot k}={\Ac}_{\cdot \cdot k_1}\otimes {\Bc}_{\cdot \cdot k_2},
\]
where 
\[
k_1= \lceil (k-1)/d_3\rceil+1,~
k_2= k-(k_1-1)d_3, 
\]
Here, $\lceil x \rceil$ stands for the largest integer no greater than $x$. 
The Sparse Kronecker Product Decomposition (SKPD) \citep{wu2023sparse} introduces a $R$-term factorization:
\begin{equation}
    \Cc= 
\sum_{r=1}^R \Ac_r \otimes \Bc_r,\label{eq: C}
\end{equation}
where $\otimes$ is the tensor Kronecker product, each $\Ac_r\in \mathbb{R}^{p_1\times p_2 \times p_3}$ identifies salient spatial or structural regions (often enforced to be sparse), and $\Bc_r\in \mathbb{R}^{d_1\times d_2 \times d_3}$ encodes the corresponding shapes or intensities. The sparsity assumptions on $\Ac$ and $\Bc$ are crucial for identifying signal regions, implying that only a few blocks of unknown shapes within the coefficient matrices contain nonzero signals. In addition, we define the transformation operator $\calT: \Real^{(p_1 d_1) \times (p_2 d_2) \times (p_3 d_3)} \mapsto  \Real^{(p_1 p_2 p_3) \times (d_1 d_2 d_3)}$ and let $\Cc^{d_1,d_2,d_3}_{k,l,m} \in \Real^{d_1\times d_2\times d_3}$ be the ($k,l,m$)-th block of $\Cc$ to represent a local structure.  

Formally, {\small
\begin{eqnarray}
\label{eq:transX}
\widetilde{\Cb} &\overset{def}{=}& \mathcal{T}(\Cc)\notag\\
   &=& \begin{array}{cccccccc}
        \bigg( \vecop(\Cc_{1,1,1}^{d_1,d_2,d_3}),
    & \dots, & \vecop(\Cc_{1,1,p_3}^{d_1,d_2,d_3}),
    &\dots,&
 \vecop(\Cc_{1,p_2,1}^{d_1,d_2,d_3}),
    & \dots, & \vecop(\Cc_{1,p_2,p_3}^{d_1,d_2,d_3}),&\dots
    \end{array}   \notag \\
   && \begin{array}{cccccccc}
    ~~~~~~~&\vecop(\Cc _{p_1,1,1}^{d_1,d_2,d_3}),
    & \dots, & \vecop(\Cc_{p_1,1,p_3}^{d_1,d_2,d_3}),
    &\dots,&
 \vecop(\Cc_{p_1,p_2,1}^{d_1,d_2,d_3}),
    & \dots, & \vecop(\Cc_{p_1,p_2,p_3}^{d_1,d_2,d_3}) \Bigg)^\top
    \end{array}.\notag\\
\end{eqnarray}}
Moreover, 
 
 \begin{equation}
\label{eq:transC}
   \calT( \Cc )= \calT(\sum^R_{r=1}\Ac_r \otimes \Bc_r)= \sum^R_{r=1}\vecop(\Ac_r)\vecop(\Bc_r)^\top. 
\end{equation}
 This transformation enables matrix-based analysis of tensor SKPD, reducing computational complexity and enhancing interpretability by separating location $\Ac_r$ and signal characteristics $\Bc_r$. 
From \eqref{eq:transX} and \eqref{eq:transC}, we have 
\begin{equation}
\langle {\cal X}_i,\Cc \rangle=
\langle {{\cal T}(\cal X}_i), {\cal T}(\Cc) \rangle=
\sum^R_{r=1} \ab^\top_r \widetilde{\Xb_i}\bb_r, \label{rterm}
\end{equation}
where $\ab_r= \vecop(\Ac_r)$ and $\bb_r= \vecop(\Bc_r)$. 
From \eqref{gaussian} and \eqref{eq: C},  Wu and Feng \citep{wu2023sparse} proposed SKPD to estimate the parameters
$\{\ab_r, \bb_r\}_{r=1}^R$ and $\gamma$. Especially when $R=1$, they consider 
\begin{eqnarray}
&&(\widehat{\Ac}, \widehat{\Bc})
\in \underset{\widehat{\Ac},~\widehat{\Bc}}{\rm argmin}\,
\frac{1}{2}
(y_i-
\langle {\calX}_i, {\Ac}\otimes {\Bc}\rangle)^2+\lambda\|\vecop(\Ac)\|_1,~\lambda>0\notag\\
&&\mbox{subject to~~}\|\Ac\|_F=1. 
\end{eqnarray}
for a linear regression model $y_i=  \langle {\cal X}_i, {\Ac}\otimes {\Bc}\rangle+\varepsilon_i, \varepsilon_i\sim N(0,1)$, a simplified model of \eqref{gaussian}. Here, the penalty term $\|\vecop(\Ac)\|_1$ is introduced to promote sparsity, while the constraint $\|\Ac\|_F = 1$ is imposed to ensure identifiability. 

Following the mixed-type multivariate Bayesian sparse variable selection with shrinkage priors (Mt-MBSP) framework of \cite{wang2025two}, the log-posterior distribution for parameters of interest can be expressed as
$\log(\text{posterior}) = \log(\text{likelihood}) + \log(\text{prior}),$
highlighting the critical role of the prior specification in determining the form and strength of regularization. In particular, the choice of $\log(\text{prior})$ governs the penalty imposed on model complexity and directly influences sparsity and shrinkage behaviour. Now, we develop a Bayesian SKPD methodology based on MtMBSP 
in order to enable a more flexible approach to traditional regularization. To apply the Gibbs sampler to binary responses used in MtMBSP, we convert the logistic model in~\eqref{logist} to a conditionally Gaussian form via P\'olya–Gamma augmentation \citep{polson2013bayesian}.  For $\omega\sim\calPG(a,b)$, the P\'olya-Gamma distribution admits the series representation
\[
\omega \;=\;\frac{1}{2\pi^{2}}\sum_{\ell=1}^{\infty}
\frac{g_{\ell}}{\bigl(\ell-\tfrac12\bigr)^{2}+(\tfrac{b}{2\pi})^{2}},
\qquad g_{\ell}\stackrel{\text{i.i.d.}}{\sim}\text{Gamma}(a,1).
\]
Polson’s identity then states
\[
\frac{e^{f_{1}\theta}}{(1+e^{\theta})^{f_{2}}}
=2^{f_{2}}e^{\kappa\theta}\int_{0}^{\infty}
e^{-\omega\theta^{2}/2}\calPG(\omega\mid f_{2},0)\,d\omega,
\quad \kappa=f_{1}-\tfrac12 f_{2}.
\]
Hence the joint density factorizes as
\begin{eqnarray*}
p(y,\omega\mid\theta)=2^{-f_{2}}
e^{\kappa\theta-\frac{\omega\theta^{2}}{2}}
\,\calPG(\omega\mid f_{2},0).
\end{eqnarray*}
Introducing the working variable $\tilde y$ and weight $\omega$,
\begin{eqnarray}
(\tilde y,\omega)=
\begin{cases}
\bigl((y-\tfrac12)/\omega,\;\omega\bigr), & y\ \text{is binary},\ \omega\sim\calPG(1,0),\\[4pt]
(y,\,1), & y\ \text{is continuous}.
\end{cases}\label{pg}
\end{eqnarray}
Both the logistic model~\eqref{logist} and the Gaussian model~\eqref{gaussian} can be expressed within a unified framework:
\[
p(\Theta\mid \tilde y,\omega,u)\;\propto\;
p(\Theta)\exp\!\Bigl\{-\frac{\omega}{2}\bigl(\tilde y-\Theta\bigr)^{2}\Bigr\},
\qquad
\Theta=\langle\mathcal{X},\mathbf{C}\rangle
      +\mathbf{z}^{\top}\boldsymbol{\gamma}+u.
\]
This Gaussian form permits straightforward Gibbs updates for
$\mathbf{C}$ and $\boldsymbol{\gamma}$ under the three-parameter beta-normal (TPBN) priors, while the
Kronecker decomposition $\Cc
 =\Ac\otimes\Bc$ (for each response $k$) captures
low-rank spatial structure across tensor predictors.

\subsection{Posterior and Algorithm}\label{sec:comp}

To formally describe the proposed Bayesian framework, we begin by introducing notation. Let $p=p_1p_2p_3$ and $d=d_1d_2d_3$. Let $\widetilde{\Yb} = (\tilde y_{ik})$ denote an $n \times K$ response matrix, and let $\Zb = (\zb_1, \dots, \zb_n)$ be a $q \times n$ matrix of covariates. Define $\Ub=(u_{ik})$ as an $n \times K$ latent matrix, and $\Omegab_k = \mathrm{diag}(\omega_{1k}, \dots, \omega_{nk})$ as an $n \times n$ diagonal weighted matrix, depended on $\tilde y$. 

The regression coefficients are stacked as $\Ab_k = (\ab_{1k}, \dots, \ab_{Rk})$ and $\Bb_k = (\bb_{1k}, \dots, \bb_{Rk})$. We define a $d \times R$ random matrix $\Bb$ following the distribution $\Bb\sim \mathrm{TPBN}{d \times R}(a, u, \tau)$, specified by
\begin{eqnarray}
\Bb\mid \zeta_1,\dots,\zeta_d&\sim&  
{\cal MN}_{d\times R}(
\Ob,~{\rm diag}(\zeta_1,\dots,\zeta_d),\Ib_R
),\notag\\
\zeta_j\mid \xi_j&\overset{ind}{\sim}&{\cal G}(u,\xi_j), \notag \\
\xi_j&\overset{iid}{\sim}&
{\cal G}(a,\tau),~~j=1,\dots,d.
\label{tpbn1}
\end{eqnarray}

From \eqref{rterm}, the posterior density for the parameters of interest takes the form 
\begin{eqnarray}
&&p(\{{\Ab}_k,{\Bb}_k,\gammab_k\}_{k=1}^K
\mid 
 \{\Omegab\}_{k=1}^K, \Ub
)\notag\\
&\propto& \prod_{k=1}^K p({\Ab}_k)p(\Bb_k)p(\gammab_k)
\exp\left\{
-\frac{1}{2}\sum^K_{k=1}\sum^n_{i=1} \omega_{ik}(\tilde y_{ik} - {\rm tr}( \Ab^\top_k \widetilde{\Xb}_i \Bb_k)
-\gammab_k^\top \zb_i - u_{ik})^2
\right\},\label{likelihood}
\end{eqnarray}
Each of the parameter blocks, $\Ab_k$, $\Bb_k$, and $\gammab_k$ is assigned a three-parameter beta-normal (TPBN) prior:
\begin{eqnarray}
\gammab_k &\sim& {\rm TPBN}_{q\times 1}(a^{[0]},u^{[0]},\tau^{[0]}),\notag\\
\Ab_k &\sim& {\rm TPBN}_{p\times R}(a^{[1]},u^{[1]},\tau^{[1]}),\notag\\
\Bb_k &\sim& {\rm TPBN}_{d\times R}(a^{[2]},u^{[2]},\tau^{[2]}),~~K=1,\dots,K,\label{tpbn2}
\end{eqnarray}
where the parameters 
$(a^{[\ell]},u^{[\ell]})$ control local shrinkage, while 
$\tau^{[\ell]} $ governs global shrinkage. In addition,  the latent matrix
$\Ub$ is assigned as below. 
\begin{eqnarray}
\Ub\sim{\cal MN}_k(\Ob_{n\times K},~\Ib_n,~ \Sigmab)\mbox{~~and~~}
\Sigmab \sim {\cal IW}(c_0,~c_1\Ib_K)\mbox{~~~~for $c_0,~c_1>0$}, 
\label{U}
\end{eqnarray}
where the covariance $\Sigmab$ has an Inverse-Wishart (${\cal IW}$) posterior, regularized to ensure numerical stability. Therefore, we derive the posterior distribution of the parameters $\{\Ab_k, \Bb_k, \gammab_k, \Omegab_k\}_{k=1}^K$, $\Ub$, and $\Sigmab$ based on equations~\eqref{pg},\eqref{tpbn1}, \eqref{tpbn2}, and~\eqref{U}, following the arguments of \cite{wang2025two}. Posterior inference is then carried out using an efficient Gibbs sampling scheme within a Markov chain Monte Carlo (MCMC) framework. Define two $nK\times nK$ matrices:  $\widetilde{\Omegab}={\rm blockdiag}(\Omegab_1,\dots,\Omegab_K)$, $\Gammab=(\gammab_1,\dots,\gammab_K)$. Let 
$\Xbbb=(
\vecop(\tilde{\Xb}_1),\vecop(\tilde{\Xb}_2),\dots, \vecop(\tilde{\Xb}_n))$, $\Cbbb=(\vecop(\Cb_1),\dots,\vecop(\Cb_K))$ and $\Ub=(\ub_1,\dots,\ub_k)$. The complete algorithm is outlined below.

\begin{algorithm}[H]
\caption{Gibbs sampler for the BSKPD model}
\label{alg:gibbs}
\KwIn{
Tensor covariates $\{{\cal X}_i\}_{i=1}^n \subset \mathbb{R}^{D_1 \times D_2 \times D_3}$, covariate matrix $\Zb \in \mathbb{R}^{q \times n}$, response matrix $\Yb \in \mathbb{R}^{n \times K}$; \\
tensor mode dimensions $(p_\ell, d_\ell)$ with $D_\ell = p_\ell d_\ell$ for $\ell=1,2,3$; Tucker rank $R$; \\
hyperparameters $\{a^{[\ell]}, u^{[\ell]}, \tau^{[\ell]}\}_{\ell=0}^2$, $c_0$, $c_1$; total iterations $T$ and burn-in $B$.
}
\KwOut{
Posterior medians $\{\widehat{\Cc}_k, \widehat{\gammab}_k\}_{k=1}^K$, and $\widehat{\Sigmab}$.
}

\textbf{Step 1.} For each $i=1,\dots,n$, apply unfolding operator ${\cal T}$ to obtain matrix covariates 
$
\widetilde{\Xb}_i = {\cal T}({\cal X}_i) \in \mathbb{R}^{p \times d}, \quad
p = p_1 p_2 p_3, \quad d = d_1 d_2 d_3.$ 

\vspace{1ex}
\textbf{Step 2.} Initialize parameters:

\For{$k=1$ \KwTo $K$}{
  Sample
$\Ab_k \sim \mathcal{MN}_{p \times R}\bigl(
  \mathbf{0}_{p \times R}.
  , 1/\sqrt{p}\Ib_p, 1/\sqrt{R} \Ib_R \bigr), \quad
  \Bb_k \sim \mathcal{MN}_{d \times R}\bigl( \mathbf{0}_{d \times R}., 1/\sqrt{d} \Ib_d, 1/\sqrt{R}\Ib_R \bigr)$,\\
  ~~$\gammab_k \sim \mathcal{N}_q\left(0, 1/\sqrt{q} \Ib_q\right);$ Set 
  $\omega_{\cdot k} \sim \mathcal{PG}(1,0), \quad
  \Sigmab \leftarrow \Ib_K, \quad \Ub \leftarrow \mathbf{0}_{n \times K}.$
}

\vspace{1ex}
\textbf{Step 3.} Run Gibbs sampling:

\For{$t=1$ \KwTo $T$}{
  \For{$k=1$ \KwTo $K$}{
    Compute $\Cb_k \leftarrow \Bb_k \Ab_k^\top.$ Form linear predictor 
    $\Theta_k = \Xbbb^\top \mathrm{vec}(\Cb_k) + \Zb^\top \gammab_k + \ub_k$,\\
    \If{response is continuous}{
      Set $\omegab_{\cdot k} \leftarrow 1$ and $\tilde{\yb}_k \leftarrow \yb_k$.
    }
    \Else{
      Sample
      \[
      \omegab_{\cdot k} \sim \mathcal{PG}(1, \Theta_{\cdot k}), \quad
      \tilde{\yb}_{\cdot k} \leftarrow \frac{\yb_{\cdot k} - 0.5}{\omegab_{\cdot k}}.
      \]
    }

    Update $(\Ab_k, \Bb_k, \gammab_k)$ from their TPBN–Gaussian conditionals.
  }

  Update latent factors:
  \[
  \mathrm{vec}(\Ub) \sim \mathcal{N}_{nK}\bigl(
    \Deltab^{-1} \widetilde{\Omegab} \mb,
    \Deltab^{-1}
  \bigr),
  \]
  where
  \[
  \mb = \vecop(\widetilde{\Yb} - \Xbbb^\top \Cbbb - \Zb^\top \Gammab),
  \quad
  \Deltab = (\Sigmab \otimes \Ib_n)^{-1} + \widetilde{\Omegab}.
  \]

  Update covariance matrix:
  \[
  \Sigmab \sim \mathcal{IW}\bigl(n + K + c_0, \Ub^\top \Ub + c_1 \Ib_K \bigr).
  \]

  \If{$t > B$}{
    Store current draws.
  }
}

\vspace{1ex}
\textbf{Step 4.} After burn-in, for each $k=1,\dots,K$, represent $\Cb_k = \Bb_k \Ab_k^\top$ and compute posterior medians of all elements of $\Cb_k$, $\gammab_k$, and $\Sigmab$. These medians are the point estimators $\widehat{\Cb}_k$, $\widehat{\gammab}_k$, and $\widehat{\Sigmab}$.

\vspace{1ex}
\textbf{Step 5.} Using inverse function ${\cal T}^{-1}$, obtain $\widehat{\Cc}_k={\cal T}^{-1}(\widehat{\Cb}_k)$.  
\end{algorithm}
~\\[.2in]

In Algorithm~\ref{alg:gibbs}, all conditional densities for $\Ab_k$, $\Bb_k$, $\gammab_k$, $\Ub$, $\omega_{ij}$, and $\Sigmab$ admit closed-form expressions. Consequently, Gibbs sampling is implemented by iteratively drawing each variable from its full conditional distribution, conditioned on the current values of all other variables. The TPBN-Gaussian full conditional distributions for $\{\Ab_k, \Bb_k, \gammab_k\}_{k=1}^K$ are detailed in the Supplementary Material. In Step 4, we calculate the posterior medians for $\mathbf{C}_k$ rather than for $\Ab_k$ and $\Bb_k$, since the conditional distributions of $\Ab_k$ and $\Bb_k$ are not unimodal due to the non-identifiability of their true values $\{\Ab_{k0}, \Bb_{k0}\}$. Further discussion on the identifiability issue is provided in the next section.

\section{Theoretical Properties}
\label{sec:theory}
In Section \ref{sec:id}, we discuss the identifiability issues, focusing particularly on $\Ab_k$, $\Bb_k$, and their composite $\Cb_k = \Bb_k \Ab_k^\top$. In Section \ref{sec:cs}, we present the relevant theoretical results along with the necessary conditions. For notational simplicity, we suppress the subscript $k$ on $\gammab_k$, $\Ab_k$, and $\Bb_k$ throughout this section.

\subsection{Identifiability issue}\label{sec:id}
Given that the coefficient matrices $\Ab$ and $\Bb$ in \eqref{likelihood} are functions of the sample size $n$, we proceed by introducing the following definition to rigorously address the issue of identifiability.  
\begin{definition}
Let $\Ab_n$ and $\Ab'_n$ be two matrices that depend on the sample size $n$. We say that $\Ab'_n$ is asymptotically identical to $\Ab_n$ as $n \to \infty$ if ${\rm liminf}\|\Ab_n - \Ab'_n\|_F/\|\Ab_n\|_F \to 0$. 
\end{definition}

\noindent 
 From  
\eqref{tpbn1} and \eqref{tpbn2},  
\begin{eqnarray}
p(\gammab \mid \Psib^{[0]}) &\propto&
\exp\left(
-\frac{1}{2}  \gammab^\top \{\Psib^{[0]}\}^{-1} \gammab\right),\notag\\
p(\Ab \mid \Psib^{[1]}) &\propto&
\exp\left(
-\frac{1}{2} {\rm tr}\left( \Ab^\top \{\Psib^{[1]}\}^{-1}
\Ab \right)\right),\notag\\
p(\Bb \mid \Psib^{[2]}) &\propto&
\exp\left(
-\frac{1}{2} {\rm tr}\left(
\Bb^\top \{\Psib^{[2]}\}^{-1}\Bb\right)\right),
\end{eqnarray}
where  $\Psib^{[0]}
= 
{\rm diag}\left(\zeta^{[0]}_1,\dots,\zeta^{[0]}_{q}\right)
$, 
$\Psib^{[1]}
= 
{\rm diag}\left(\zeta^{[1]}_1,\dots,\zeta^{[1]}_{p}\right)
$, and $\Psib^{[2]}
= 
{\rm diag}\left(\zeta^{[2]}_1,\dots,\zeta^{[2]}_{d}\right). 
$

\noindent 
Based on equation \eqref{likelihood}, the posterior probability can be expressed as 
\begin{eqnarray}
&&\log\,p(\{\Ab_k,\Bb_k,\gamma_k\}_{k=1}^K\mid 
\{\Psib^{[\ell]},\}_{\ell=0}^2, \{\Omegab_k\}_{k=1}^K, \Ub
)\notag\\
&=&-
\frac{1}{2}\sum^K_{k=1} \left(\tilde \yb_{k} - \Xb^\top 
\vecop(\Bb_k\Ab_k^\top)
- \Zb^\top \gammab_k -\ub_k\right)^\top \Omega_k
\left(\tilde \yb_{k} - \Xb^\top 
\vecop(\Bb_k\Ab_k^\top)
- \Zb^\top \gammab_k -\ub_k\right)\notag\\
&&-\frac{1}{2}
 \gammab^\top_k \{\Psib^{[0]}\}^{-1} \gammab_k
-\frac{1}{2} {\rm tr}\left( \Ab_k^\top \{\Psib^{[1]}\}^{-1}
\Ab_k \right)
-\frac{1}{2} {\rm tr}\left( \Bb_k^\top \{\Psib^{[2]}\}^{-1}
\Bb_k \right)+{\rm Rermainder
},
\label{qq}\notag
\end{eqnarray}
where Remainder is independent of  
$\{\Ab_k,\Bb_k\}_{k=1}^K$ and $\gammab$. 
Define 
\[ L(\Ab,\Bb,\gammab)= \frac{1}{n}
{\rm E}[\log\,p(
\Ab,\Bb,\gamma \mid 
\{\Psib^{[\ell]}\}_{\ell=0}^2, \Omegab, \Ub
)]
\]
and \begin{eqnarray}
Q(\Cb,\gammab)
=\frac{-1}{n}{\rm E}\left[\left(\tilde \yb - \Xb^\top 
\vecop(\Cb)
- \Zb^\top \gammab -\ub\right)^\top \Omega
\left(\tilde \yb - \Xb^\top 
\vecop(\Cb)
- \Zb^\top \gammab -\ub\right)\right].\label{QC}
\end{eqnarray}
Then 
\begin{eqnarray*}
L(\Ab,\Bb,\gammab)&=& Q(\Bb\Ab^\top,\gammab)-\frac{1}{2n}
 \gammab^\top {\rm E}[\{\Psib^{[0]}\}^{-1}]\gammab\notag\\
&&-\frac{1}{2n} {\rm tr}\left( \Ab^\top {\rm E}[\{\Psib^{[1]}\}^{-1}]
\Ab\right)
-\frac{1}{2n} {\rm tr}\left( \Bb^\top {\rm E}[\{\Psib^{[2]}\}^{-1}]
\Bb \right)+{\rm Constant.
}
\end{eqnarray*}
It can be observed that ${\rm E}[\{\Psib^{[0]}\}^{-1}]$, ${\rm E}[\{\Psib^{[1]}\}^{-1}]$, and ${\rm E}[\{\Psib^{[2]}\}^{-1}]$ are identical matrices up to a scalar multiple. Therefore, we let 
${\rm E}[\{\Psib^{[0]}\}^{-1}]=\psi_0\Ib_q$, 
${\rm E}[\{\Psib^{[1]}\}^{-1}]=\psi_1\Ib_p$, and ${\rm E}[\{\Psib^{[2]}\}^{-1}]=\psi_2\Ib_d$, where $\psi_0, \psi_1,\psi_2>0$. 
Then we have the following theorem. 

\begin{theorem}\label{thm_id}
Suppose that the function $Q(\Cb, \gammab)$ in \eqref{QC} admits a unique maximizer $\{\Cb_0, \gammab_0\}$. Assume that 
$ Q(\Cb_0, \gammab_0)>
\limsup_{n \rightarrow \infty} Q(\Cb_n, \gammab_n)$ for any sequence $\{\Cb_n, \gammab_n\}$ that is not asymptotically identical to $\{\Cb_0, \gammab_0\}$ as $n \to \infty$. Define the parameter space
\[
\Theta_n = \left\{ (\Ab_n, \Bb_n, \gammab_n) \,\mid~~
\psi_0\|\gammab_n\|^2,~~\psi_1\|\Ab_n\|_F^2,\mbox{~and~}\psi_2\|\Bb_n\|^2_F \mbox{~are all of order $o(n)$~}\right\}.
\]
Then any maximizer of $L(\Ab, \Bb, \gammab)$ within $\Theta_n$ is asymptotically identical to $(\Ab_0, \Bb_0, \gammab_0)$, where $(\Ab_0, \Bb_0)$ is determined by the following constraints. 
\begin{eqnarray}
\Cb_0 = \Bb_0 \Ab_0^\top,~\Bb_0\Bb^\top_0
=\frac{\psi_1}{\psi_2}(\Cb_0\Cb^\top_0)^{1/2},\mbox{~~and~~} \Ab_0\Ab^\top_0 
=\frac{\psi_2}{\psi_1}(\Cb^\top_0 \Cb_0)^{1/2}. \label{cc}
\end{eqnarray}
\end{theorem}



From Theorem \ref{thm_id}, the matrices $\Ab_0$ and $\Bb_0$ are not uniquely determined; indeed, for any orthogonal matrix $\Gb$, the transformed matrices $\Ab_0' = \Ab_0 \Gb$ and $\Bb_0' = \Bb_0 \Gb$ also satisfy the constraints in \eqref{cc}. Nevertheless, the parameter of interest, $\Cb_0$, is uniquely identified, and thus it suffices to obtain any single admissible pair $(\Ab_0, \Bb_0)$ in order to compute $\Cb_0$.

\subsection{Posterior consistency}\label{sec:cs} 
This section establishes the asymptotic properties of the proposed estimators $\widehat{\Cb}$ and $\widehat{\gammab}$. We begin by introducing the required conditions, which are grouped into three categories: Condition~(L), Condition~(H), and a set of regularity conditions. Condition~(L) corresponds to the classical low-dimensional setting, while Condition~(H) pertains to the high-dimensional regime. The regularity conditions are required to ensure appropriate behavior of the TPBN priors. These conditions 
 are similar to Assumptions in \cite{bai2018high}. 
Let the function $Q(\Cb, \gammab)$ defined in~\eqref{QC} admit a unique maximizer $\{\Cb_0, \gammab_0\}$ under the true models~\eqref{logist} and~\eqref{gaussian}, where the true parameter $\Cb_0$ admits a factorization of the form $\Cb_0 = \Bb_0 \Ab_0^\top$ for some matrices $\Ab_0$ and $\Bb_0$.

Let $S = S^{[1]} \times S^{[2]}$ denote the set of associated row-column index pairs. For each $i = 1, \dots, n$, let $\widetilde{\Xb}^S_i$ denote the submatrix of $\widetilde{\Xb}_i$ corresponding to the rows indexed by $S^{[1]} \subset \{1, \dots, p\}$ and the columns indexed by $S^{[2]} \subset \{1, \dots, d\}$.  Let $S' \subset \{1, \dots, q\}$ be a set of indices, and let $\Zb^{S'}$ denote the submatrix of $\Zb$ consisting of the columns indexed by $S'$.
Let $S_0 = S^{[1]}_0 \times S^{[2]}_0$ denote the index set of nonzero elements in the true parameter $\Cb_0$, where $S^{[1]}_0 \subset \{1, \dots, p\}$ indexes the nonzero rows of $\Ab_0$ and $S^{[2]}_0 \subset \{1, \dots, d\}$ indexes the nonzero rows of $\Bb_0$. Similarly, let $S'_0 \subset \{1, \dots, q\}$ denote the index set of nonzero components of $\gammab_0$. \\[.2in]
 
\noindent 
\textbf{Condition (L)}
\begin{itemize}
\item[] (L1) $ pd =o(n)$ for all $n\in {\mathbb N}$, where $p=p_1p_2p_3$ and $d=d_1d_2d_3$. 
\item[] (L2)  There exits a finite constant $\sigma>0$ so that 
\[
\frac{1}{\sigma}\leq 
{\rm liminf}_n\lambda_{\min}\left(n^{-1}\sum_{i}^n\vecop(\widetilde{\Xb}_i)\vecop(\widetilde{\Xb}_i)^\top\right)\leq {\rm limsup}_n\lambda_{\max}\left(n^{-1}\sum^n_{i=1}\vecop(\widetilde{\Xb}_i)\vecop(\widetilde{\Xb}_i)^\top\right)\leq \sigma. 
\] 
\item[] (L'1) $q =o(n)$ for all $n\in {\mathbb N}$. 
\item[] (L'2)  There exits a finite constant $\sigma'>0$ so that 
\[
\frac{1}{\sigma'}\leq 
{\rm liminf}_n\lambda_{\min}\left(n^{-1}\Zb\Zb^\top\right)\leq {\rm limsup}_n\lambda_{\max}\left(n^{-1}\Zb\Zb^\top\right)\leq \sigma'. 
\] 
\end{itemize}
Condition~(L1), which assumes $pd = O(n)$, implies that both $p = o(n)$ and $d = o(n)$. Conditions~(L2) and~(L'2) specify regularity assumptions on the sample covariance matrix of the covariates.\\[.2in]

\noindent 
\textbf{Condition (H)}
\begin{itemize}
\item[] (H1) $ n < pd$ for all $n\in {\mathbb N}$, and  $\log(pd) = O(n^\alpha)$ for some $ \alpha \in (0, 1) $. 
\item[] (H2) For any such set $S$, $|S|<n$, there exists a finite constant $\sigma>0$ so that 
\[
\frac{1}{\sigma}\leq 
{\rm liminf}_n\lambda_{\min}\left(n^{-1}\sum_{i}^n\vecop(\widetilde{\Xb}^S_i)\vecop(\widetilde{\Xb}^S_i)^\top\right)\leq {\rm limsup}_n\lambda_{\max}\left(n^{-1}\sum^n_{i=1}\vecop(\widetilde{\Xb}_i)\vecop(\widetilde{\Xb}_i)^\top\right)\leq \sigma. 
\]
\item[](H3) Let  
$S^{[1]}_0$ and $S^{[2]}_0$ are respectively 
nonempty sets of indices and $|S_0|=o(n/\log\,(pd))$.  
\item[] (H'1) $ n < q$ for all $n\in {\mathbb N}$, and  $\log\,q = o(n^{\alpha'})$ for some $ \alpha' \in (0, 1)$. 
\item[] (H'2) For any such set $S'$, $|S'|<n$,  there exists a finite constant $\sigma'>0$ so that 
\[
\frac{1}{\sigma'}\leq 
{\rm liminf}_n\lambda_{\min}\left(n^{-1}\Zb^{S'}(\Zb^{S'})^\top\right)\leq {\rm limsup}_n\lambda_{\max}\left(n^{-1}\Zb^{S'}(\Zb^{S'})^\top\right)\leq \sigma'. 
\]
\item[](H'3) Let $S'_0$ be nonempty set of indices in the true model and $|S'_0|=o(n/\log\,(q))$.
\end{itemize}

Condition~(H1) allows either $p$ or $q$ to be smaller than $n$, while both $\log(p)$ and $\log(q)$ are of order $n^\alpha$ for some $\alpha \in (0,1)$. Conditions~(H2) and~(H'2) impose regularity assumptions on the sample covariance matrix of the covariates in the high-dimensional setting. Conditions~(H3) and~(H'3) specify sparsity assumptions on the true parameters $\Ab_0$, $\Bb_0$, and $\gammab_0$, respectively. In particular, $S_0$ denotes the set of index pairs corresponding to the nonzero components of $\Ab_0$ and $\Bb_0$.\\[.2in]

\noindent 
\textbf{Regularity conditions}
\begin{itemize}
\item[](C1) The components of $\Ab_0, \Bb_0$ and $\gammab_0$ in the true model are uniformly bounded above and below by positive constants. That is, 
$0<\max\{ \sup_{j,k}|(\Ab_0)_{jk}|,~\sup_{j,k}|(\Bb_0)_{jk}|,\sup_j|(\gammab)_j|\}<\infty$. 
\item[](C2) $\lambda_{\max}(\Ab_0\Ab_0^\top)$ and 
$\lambda_{\max}(\Bb_0\Bb_0^\top)$ are nonzero and bounded. 
\item[](C3) For TPBN priors, $a^{[\ell]},u^{[\ell]}\in (0,1)$ are fixed positive numbers and $\tau^{[\ell]}\in (0,1)$, $\ell=0,1,2$. Moreover, $\tau^{[0]}=o(q^{-1}n^{-\rho_0})$, $\tau^{[1]}=o(p^{-1}n^{-\rho_1})$, $\tau^{[2]}=o(d^{-1}n^{-\rho_2})$ for $\rho_0,\rho_1,\rho_2>0$. 
\end{itemize}

Conditions~(C1) and~(C2) impose boundedness on the true parameters $\Ab_0$, $\Bb_0$, and $\gammab_0$, ensuring that the MCMC algorithm can effectively explore the posterior distribution and concentrate around the true values. Condition~(C3) specifies a suitable decay rate for the global shrinkage parameter~$\tau$. We are now in a position to present the main theorem establishing posterior consistency of the proposed method.

\begin{theorem} \label{thm:main} Assume that 
Condition (L) or Condition (H) holds. Assume that 
Conditions (C1)-(C3) hold. Suppose that 
${\rm tr}(\Sigmab)$ is positive and finite. 
Then for any $ \varepsilon > 0 $, as $n\rightarrow\infty$ 
\begin{eqnarray*}
\Pr(\Cb_n : \|\Cb_n - \Cb_{0}\|_F < \varepsilon \mid \Yb_n) \to 0 \text{~a.s.} \mbox{~~~and~~~}
\Pr(\gammab : \|\gammab_n - \gammab_0\| < \varepsilon \mid \Yb_n) \to 0\text{~a.s.} 
\end{eqnarray*}
\end{theorem}

\section{Experiments}\label{sec:experiments}

\subsection{Simulation Data Generation}
\label{sec:sim-fitting}

We simulate $ n = 400 $ samples, each with a matrix predictor $ X_i \in \mathbb{R}^{64 \times 64} $, and partition the data into a training set ($ n_{\mathrm{tr}} = 200 $) and a test set ($ n_{\mathrm{ts}} = 200 $). The true coefficient matrix $ \Cc_{\mathrm{true}} $ is a $ 64 \times 64 $ grayscale butterfly image, normalized to the range $[0, 1]$. For half of the samples, the predictors $ X_i $ are generated as $ \Cc_{\mathrm{true}} $ corrupted with Gaussian noise (with standard deviation equal to that of the butterfly image, $ \sigma = 0.2682 $); the remaining samples consist of the Gaussian noise. The responses are generated according to the following models:
\[
Y_{i1} = \mathrm{tr}({\cal X}_i^\top \Cc_{\mathrm{true}}) + \epsilon_i, \quad \epsilon_i \sim \mathcal{N}(0, 0.1),
\]
\[
\mathbb{P}(Y_{i2} = 1) = \frac{1}{1 + \exp\left\{-\mathrm{tr}({\cal X}_i^\top \Cc_{\mathrm{true}})\right\}}.
\]

\begin{figure}[H]
  \centering
  \includegraphics[width=0.9\textwidth]{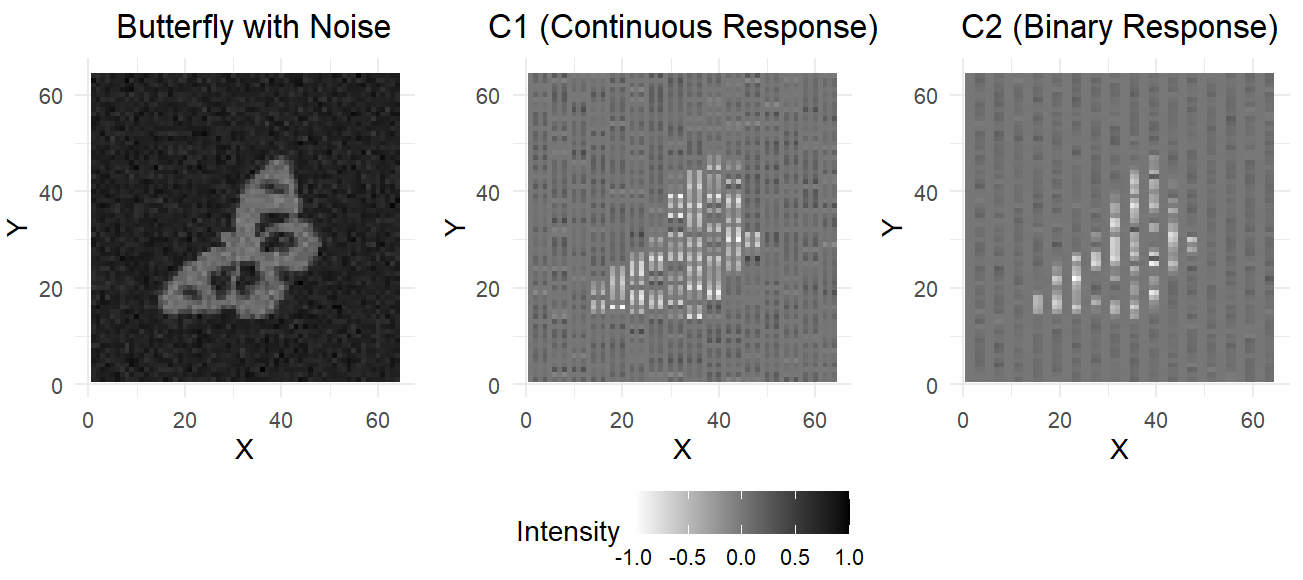}
  \caption{Noisy predictor image ${\cal X}_i$ (left) and posterior median estimates of
  the regression coefficients for the continuous ($\hat \Cc_{1}$, centre) and
  binary ($\hat \Cc_{2}$, right) responses.  All panels share the intensity
  range $[-1,1]$.}
  \label{fig:coeff}
\end{figure}

We ran the Markov chain Monte Carlo algorithm described in Section~\ref{sec:comp} for 1,000 iterations, discarding the first 500 as burn-in. Predictive performance for the continuous outcome was assessed using root-mean-squared error (RMSE)
$
\mathrm{RMSE} = \left\{ n_{\mathrm{ts}}^{-1} \sum_{i \in \mathcal{T}} (\hat{y}_{i1} - y_{i1})^2 \right\}^{1/2},
$
where $ \mathcal{T} $ denotes the test set. Classification performance for the binary outcome was evaluated using the area under the receiver operating characteristic curve (AUC). The resulting performance metrics were $ \mathrm{RMSE} = 5.77 \times 10^2 $ and $ \mathrm{AUC} = 0.9230 $, indicating accurate recovery of the underlying signal across both tasks. The data-generating signal $ \Cc_{\mathrm{true}} $ corresponds to a butterfly-shaped grayscale image. Figure~\ref{fig:coeff} compares one representative noisy predictor matrix ${\cal X}_i$ to the estimated coefficient matrices $ \hat{\Cc}_1 $ (for the continuous response) and $ \hat{\Cc}_2 $ (for the binary response), computed as posterior means. All images are displayed on a common intensity scale of $[-1, 1]$, with a blue--white--red color map for the coefficients and grayscale for the predictor. Both estimates successfully recover the butterfly structure, with $ \hat{\Cc}_1 $ exhibiting slightly sharper wing contours.

These results confirm that the BSKPD model performs precise voxel
selection, retains fine spatial detail and handles mixed response types
robustly, as evidenced by the low RMSE, high AUC and coherent coefficient
maps.

\subsection{OASIS--1 and ADNI--1 studies}\label{sec:oasis_adni}

Two publicly available T1--weighted magnetic resonance imaging (MRI) datasets were analysed. The \textit{Open Access Series of Imaging Studies} (OASIS--1) comprises 416 atlas--registered brain volumes of dimension $176\times208\times176$, from individuals aged 18 to 96 years. Among them, 100 subjects aged over 60 satisfy the diagnostic criteria for very mild to mild Alzheimer’s disease (AD); see \cite{marcus2007open}. Each subject is accompanied by cognitive assessments, including the Mini–Mental State Examination (MMSE, continuous) and the Clinical Dementia Rating (CDR, ordinal, binarised at CDR$>0$). The Alzheimer’s Disease Neuroimaging Initiative (ADNI--1) ``Complete 1Yr 1.5T'' Siemens subset contains 246 participants: 77 cognitively normal (CN), 125 with mild cognitive impairment (MCI), and 44 with Alzheimer’s disease. Each volume has dimension $192\times192\times160$. Binary responses were defined as AD vs. non-AD ($y_1$) and MCI vs. CN+AD ($y_2$). All images were downsampled via nearest-neighbour interpolation to $64^3$ voxels and intensity-normalised to have zero mean and unit variance. Demographic covariates included standardised age and a sex indicator ($0$ = female, $1$ = male). One corrupted OASIS scan was excluded, yielding $n_{\textsc{oasis}} = 415$.

For subject $i$, let $\mathcal{X}_i \in \mathbb{R}^{64\times64\times64}$ denote the 3D tensor predictor, and let $\mathbf{y}_i = (y_{i1}, y_{i2})^\top$ denote the vector of responses (continuous + binary for OASIS; binary + binary for ADNI). The proposed BSKPD model assumes
\[
g_k\left\{ \mathbb{E}(y_{ik} \mid \mathcal{X}_i, \mathbf{z}_i) \right\} = \langle \mathcal{X}_i, \Cc_k \rangle + \mathbf{z}_i^\top \boldsymbol{\gamma}_k,
\quad \text{with } \Cc_k = \Ac_k \otimes \Bc_k, \quad k = 1, 2,
\]
where $g_1$ is the identity link (for MMSE) and $g_2$ the logit link (for binary outcomes), and $\mathbf{z}_i = (\text{age}, \text{sex})^\top$. Hyperparameters were set to $p_j=32$, $d_j=2$ for $j = 1,2,3$. A blocked Gibbs sampler was run for 2,000 iterations, discarding the first 1,000 as burn-in. Five-fold cross-validation was used to evaluate predictive performance. For continuous outcomes, mean squared error (MSE) was used; for binary outcomes, the area under the receiver operating characteristic curve (AUC) was computed.

\begin{figure}[ht]
    \centering
    \includegraphics[scale=0.35]{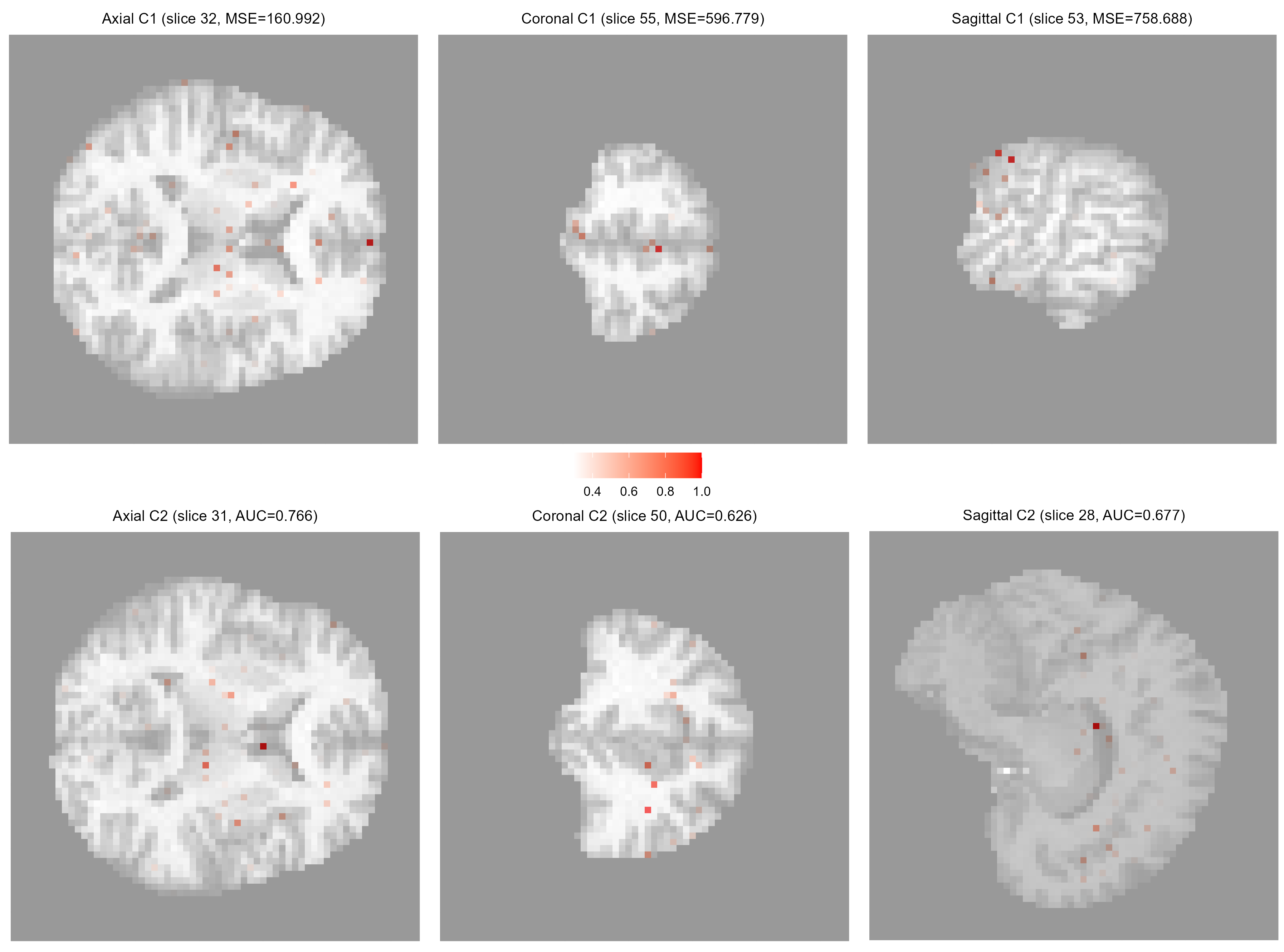}
    \caption{Significant slices of regression coefficients $\hat \Cc_1 $ (top, MMSE) and $ \hat \Cc_2 $ (bottom, CDR) overlaid on MRI background, selected by min test MSE and max test AUC. Positive elements of the coefficients range from $0$ to $ 1.0 $ in a red scale.}
    \label{fig:coeff_slices_Oasis}
\end{figure}

\begin{figure}[ht]
    \centering
    \includegraphics[width=0.95\textwidth]{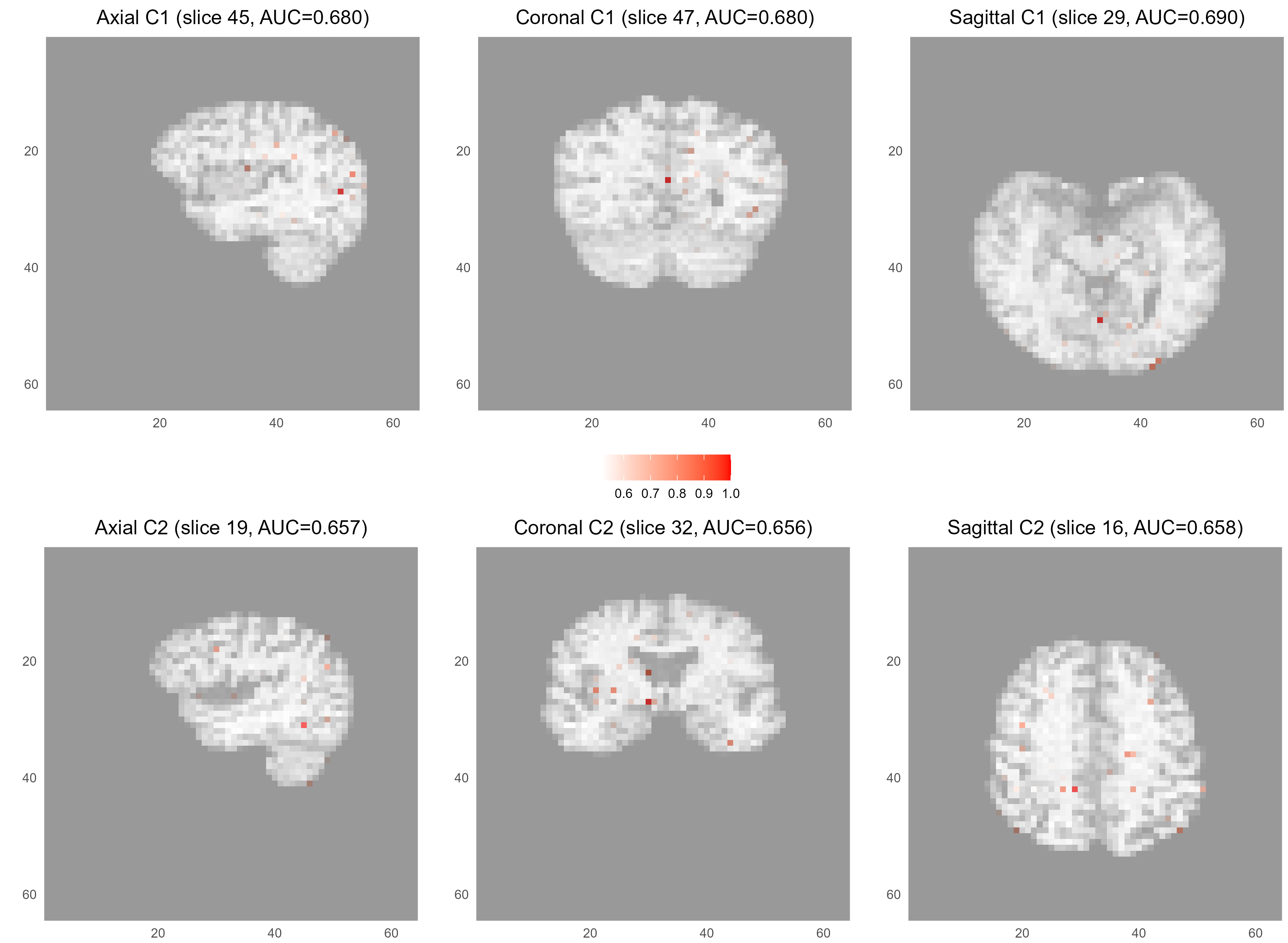}
    \caption{Significant slices of regression coefficients $\hat \Cc_1 $ (top, AD) and $\hat \Cc_2 $ (bottom,MCI) overlaid on MRI background, selected by maximum absolute value of the 5-fold cross validation MSE and AUC. Positive elements of the coefficients range from $0$ to $ 1.0 $ in a red scale.}
    \label{fig:coeff_slices_ADNI}
\end{figure}

On the OASIS dataset, the mean test-fold MSE for MMSE prediction was $8.4031 \pm 0.8990$, and the AUC for CDR classification was $0.8620 \pm 0.0131$. On the ADNI dataset, the model attained AUCs of $0.660 \pm 0.050$ for AD classification and $0.6531 \pm 0.059$ for MCI classification.
Figure~\ref{fig:coeff_slices_Oasis} displays the absolute values of the estimated coefficient tensors $\Cc_1$ (MMSE) and $\Cc_2$ (CDR), overlaid on an axial MRI slice. MMSE weights are concentrated in neocortical association areas and the hippocampal formation, regions known to underlie global cognitive function. In contrast, the CDR map emphasises medial temporal and ventricular regions associated with dementia progression. Analogous coefficient maps for ADNI (Figure~\ref{fig:coeff_slices_ADNI}) again highlight the hippocampus and surrounding medial temporal lobe, consistent with early atrophy patterns in AD and MCI \citep{mcevoy2010quantitative}. Hippocampal volume is typically reduced by 15--30\% at the MCI stage, and its accelerated decline is predictive of subsequent cognitive deterioration. The differing estimated ranks across tasks reflect variation in the spatial scale of disease-specific effects.

\section{Discussion}\label{sec:discussion}

The Bayesian Sparse Kronecker Product Decomposition with multi–response Beta–Normal shrinkage (BSKPD) provides a coherent solution to mixed-type tensor regression.  By factorising each coefficient array into a low-rank Kronecker product, it reduces the parameter count from $(D_1D_2D_3)$ to $O(p_{1}d_{1}+p_{2}d_{2}+p_{3}d_{3})$, enabling inference for ultrahigh-dimensional three-way images.  The elementwise three-parameter Beta–Normal prior, governed by a global $\mathrm{Ga}(0.3,1.5)$ scale, enforces adaptive sparsity and delivers response-specific voxel subsets with full posterior uncertainty quantification. Empirical analyses of the OASIS and ADNI cohorts corroborate these theoretical advantages.  BSKPD matches or surpasses leading black-box predictors while recovering anatomically interpretable regions.  Distinct Kronecker ranks were selected for Alzheimer’s disease ($16^{3}\!\times\!4^{3}$) and mild cognitive impairment ($32^{3}\!\times\!2^{3}$), indicating the model’s ability to adapt spatial resolution to disease stage.  Posterior credible maps highlight neocortical atrophy for AD and hippocampal–parahippocampal shrinkage for prodromal MCI, echoing multiscale findings of \cite{hett2021multi}.

Limitations include a fixed global shrinkage parameter and the assumption of independent subjects.  Future work will explore (i) hierarchical global–local priors (e.g.\ Dirichlet–Laplace) to stabilise selection across scanners, (ii) longitudinal extensions that incorporate repeated images, and (iii) semiparametric monotone links for ordinal counts and bounded scores.  With these extensions, BSKPD can serve as a principled, interpretable template for mixed-type regression across diverse high-order biomedical datasets.

\section*{Acknowledgment}
Hsin-Hsiung Huang’s work was partially supported by the National Science Foundation under grants DMS-1924792 and DMS-2318925 and 
Shao-Hsuan Wang’s research was partially supported by the Ministry of Science and Technology, Taiwan, under Grant MOST 112-2628-M-008-001-MY3. The authors are grateful to the Editor, Associate Editor and the anonymous referees for their insightful comments, which have substantially improved the manuscript.

\section*{Supplementary material}\label{SM}
The Supplementary Material available at \textit{Biometrika} online
contains the full technical appendix: complete proofs of Theorems 1–3,
supporting lemmas, additional simulation results, extra figures
illustrating voxel selection, and well-documented \texttt{R} code for
reproducing the BSKPD analyses reported in Sections \ref{sec:experiments}
and~\ref{sec:discussion}.  The paper can be read independently of this
material.

\bibliographystyle{plainnat}
\bibliography{BSKPD_arxiv}  

\begin{thebibliography}{17}
\providecommand{\natexlab}[1]{#1}
\providecommand{\url}[1]{\texttt{#1}}
\expandafter\ifx\csname urlstyle\endcsname\relax
  \providecommand{\doi}[1]{doi: #1}\else
  \providecommand{\doi}{doi: \begingroup \urlstyle{rm}\Url}\fi

\bibitem[Armagan et~al.(2011)Armagan, Clyde, and
  Dunson]{armagan2011generalized}
Artin Armagan, Merlise Clyde, and David Dunson.
\newblock Generalized beta mixtures of gaussians.
\newblock \emph{Advances in neural information processing systems}, 24, 2011.

\bibitem[Bai and Ghosh(2018)]{bai2018high}
Ray Bai and Malay Ghosh.
\newblock High-dimensional multivariate posterior consistency under
  global--local shrinkage priors.
\newblock \emph{Journal of Multivariate Analysis}, 167:\penalty0 157--170,
  2018.

\bibitem[Bhattacharya et~al.(2015)Bhattacharya, Pati, Pillai, and
  Dunson]{bhattacharya2015dirichlet}
Anirban Bhattacharya, Debdeep Pati, Natesh~S Pillai, and David~B Dunson.
\newblock Dirichlet--laplace priors for optimal shrinkage.
\newblock \emph{Journal of the American Statistical Association}, 110\penalty0
  (512):\penalty0 1479--1490, 2015.

\bibitem[Carvalho(2010)]{carvalho2010horseshoe}
CM~Carvalho.
\newblock The horseshoe estimator for sparse signal.
\newblock \emph{Biometrika}, 1:\penalty0 1, 2010.

\bibitem[George and McCulloch(1993)]{george1993variable}
Edward~I George and Robert~E McCulloch.
\newblock Variable selection via gibbs sampling.
\newblock \emph{Journal of the American Statistical Association}, 88\penalty0
  (423):\penalty0 881--889, 1993.

\bibitem[Guhaniyogi et~al.(2017)Guhaniyogi, Qamar, and
  Dunson]{guhaniyogi2017bayesian}
Rajarshi Guhaniyogi, Shaan Qamar, and David~B Dunson.
\newblock Bayesian tensor regression.
\newblock \emph{Journal of Machine Learning Research}, 18\penalty0
  (79):\penalty0 1--31, 2017.

\bibitem[Hett et~al.(2021)Hett, Ta, Oguz, Manj{\'o}n, Coup{\'e}, Initiative,
  et~al.]{hett2021multi}
Kilian Hett, Vinh-Thong Ta, Ipek Oguz, Jos{\'e}~V Manj{\'o}n, Pierrick
  Coup{\'e}, Alzheimer’s Disease~Neuroimaging Initiative, et~al.
\newblock Multi-scale graph-based grading for alzheimer’s disease prediction.
\newblock \emph{Medical image analysis}, 67:\penalty0 101850, 2021.

\bibitem[Jack~Jr et~al.(2008)Jack~Jr, Bernstein, Fox, Thompson, Alexander,
  Harvey, Borowski, Britson, L.~Whitwell, Ward, et~al.]{jack2008alzheimer}
Clifford~R Jack~Jr, Matt~A Bernstein, Nick~C Fox, Paul Thompson, Gene
  Alexander, Danielle Harvey, Bret Borowski, Paula~J Britson, Jennifer
  L.~Whitwell, Chadwick Ward, et~al.
\newblock The alzheimer's disease neuroimaging initiative (adni): Mri methods.
\newblock \emph{Journal of Magnetic Resonance Imaging: An Official Journal of
  the International Society for Magnetic Resonance in Medicine}, 27\penalty0
  (4):\penalty0 685--691, 2008.

\bibitem[Marcus et~al.(2007)Marcus, Wang, Parker, Csernansky, Morris, and
  Buckner]{marcus2007open}
Daniel~S Marcus, Tracy~H Wang, Jamie Parker, John~G Csernansky, John~C Morris,
  and Randy~L Buckner.
\newblock Open access series of imaging studies (oasis): cross-sectional mri
  data in young, middle aged, nondemented, and demented older adults.
\newblock \emph{Journal of cognitive neuroscience}, 19\penalty0 (9):\penalty0
  1498--1507, 2007.

\bibitem[McEvoy and Brewer(2010)]{mcevoy2010quantitative}
Linda~K McEvoy and James~B Brewer.
\newblock Quantitative structural mri for early detection of alzheimer’s
  disease.
\newblock \emph{Expert review of neurotherapeutics}, 10\penalty0 (11):\penalty0
  1675--1688, 2010.

\bibitem[Mitchell and Beauchamp(1988)]{mitchell1988bayesian}
Toby~J Mitchell and John~J Beauchamp.
\newblock Bayesian variable selection in linear regression.
\newblock \emph{Journal of the American Statistical Association}, 83\penalty0
  (404):\penalty0 1023--1032, 1988.

\bibitem[Polson and Scott(2010)]{polson2010shrink}
Nicholas~G Polson and James~G Scott.
\newblock Shrink globally, act locally: Sparse bayesian regularization and
  prediction.
\newblock \emph{Bayesian statistics}, 9\penalty0 (501-538):\penalty0 105, 2010.

\bibitem[Polson et~al.(2013)Polson, Scott, and Windle]{polson2013bayesian}
Nicholas~G Polson, James~G Scott, and Jesse Windle.
\newblock Bayesian inference for logistic models using p{\'o}lya--gamma latent
  variables.
\newblock \emph{Journal of the American Statistical Association}, 108\penalty0
  (504):\penalty0 1339--1349, 2013.

\bibitem[Wang and Xu(2024)]{wang2024bayesian}
Kunbo Wang and Yanxun Xu.
\newblock Bayesian tensor-on-tensor regression with efficient computation.
\newblock \emph{Statistics and its interface}, 17\penalty0 (2):\penalty0 199,
  2024.

\bibitem[Wang et~al.(2025)Wang, Bai, and Huang]{wang2025two}
Shao-Hsuan Wang, Ray Bai, and Hsin-Hsiung Huang.
\newblock Two-step mixed-type multivariate bayesian sparse variable selection
  with shrinkage priors.
\newblock \emph{Electronic Journal of Statistics}, 19\penalty0 (1):\penalty0
  397--457, 2025.

\bibitem[Wu and Feng(2023)]{wu2023sparse}
Sanyou Wu and Long Feng.
\newblock Sparse kronecker product decomposition: a general framework of signal
  region detection in image regression.
\newblock \emph{Journal of the Royal Statistical Society Series B: Statistical
  Methodology}, 85\penalty0 (3):\penalty0 783--809, 2023.

\bibitem[Zhou et~al.(2013)Zhou, Li, and Zhu]{zhou2013tensor}
Hua Zhou, Lexin Li, and Hongtu Zhu.
\newblock Tensor regression with applications in neuroimaging data analysis.
\newblock \emph{Journal of the American Statistical Association}, 108\penalty0
  (502):\penalty0 540--552, 2013.

\end{thebibliography}

\end{document}